\documentclass[letterpaper, 10pt, conference]{ieeeconf}
\IEEEoverridecommandlockouts	

\usepackage{mathtools}
\usepackage{ifpdf}
\usepackage{lipsum}
\usepackage{cite}
\usepackage{color}
\usepackage{graphicx}
\usepackage{lipsum}
\usepackage{amsmath}
\usepackage{listings}
\usepackage{xcolor}
\usepackage[linesnumbered,ruled]{algorithm2e}
\usepackage{algpseudocode}
\usepackage{float}
\usepackage{textcomp}
\usepackage{array}
\usepackage{placeins}
\usepackage{stfloats}
\usepackage{url}
\usepackage{lipsum}
\usepackage{multirow}
\usepackage{soul}
\usepackage[utf8]{inputenc}
\let\labelindent\relax
\usepackage{enumitem}
\usepackage{algorithmicx}
\graphicspath{{./Figures/}}
\usepackage{glossaries}
\usepackage{diagbox}

\title{\LARGE \bf Semi-Autonomous Electric Vehicles in Platooning Mode and Their Effects on Travel Time: A Framework for Simulation Evaluation}

\author{Aso Validi, \emph{Student Member, IEEE,}  
Nikita Smirnov, \emph{Student Member, IEEE,} and\\
Cristina Olaverri-Monreal \emph{Senior Member, IEEE}
\thanks{Johannes Kepler University Linz, Austria; Chair  Sustainable Transport Logistics 4.0. \texttt{\{aso.validi, nikita.smirnov, cristina.olaverri-monreal\}@jku.at}}}

\begin{document}
	
\maketitle

%
%

\begin{abstract}

Connected and Automated Vehicles (CAVs) have received a lot of attention in recent years. However, there are still numerous challenges in this field. In this paper, we investigated the effects of dynamic-flexible platooning on travel time by considering real-world trips data. For this purpose we extended the platooning capabilities of the 3DCoAutosim simulation platform, and proposed a dynamic-flexible model that we validated by creating use cases on traffic efficiency. We studied our dynamic-flexible platooning case for three electric vans with an autonomous leader, a semi-autonomous first follower with a driver, and an autonomous last follower. Results showed that the model developed in this study is efficient to investigate the effects of dynamic-flexible platooning on travel time. 
\end{abstract}


\section{Introduction}
\label{sec:introduction}

Connected vehicles with a high degree of automation can contribute to climate-neutrality ensuring less environmental impact and greater road safety~\cite{olaverri2020promoting}. The formation of platoons by partially automated vehicles contributes to the effective solutions that Intelligent Transport Systems (ITS) offer for future traffic efficiency~\cite{validi2020environmental, validi2021simulation}. 
One of the biggest challenges in this domain is managing the platoon after possible interruptions and reforming the platoon after a break up. When platoons fragment and reform it is referred to as dynamic-flexible platooning~\cite{hardes2019dynamic, schindler2018dynamic}.


The 3DCoAutoSim simulation platform was created with the goal of simulating V2V communication~\cite{michaeler20173d} among other capabilities. The platform has been continuously expanded to include automated functions and simulate cooperative Advanced Driving Assistance Systems (ADAS)~\cite{Olaverri-Monreal2018AutomatedAutomation, Capalar2018HypovigilanceWorkload, Olaverri-Monreal2018EffectAvoidance}. It is capable of testing applications that are also based on Vehicle-to-Infrastructure (V2I) or Vehicle-to-Pedestrian (V2P) communication by using vehicular data connections between various simulators~\cite{hussein20183dcoautosim, olaverri2018implementation, olaverri2018connection, artal2019}. It is also coupled with the Robot Operating System (ROS)~\cite{quigley2009ros} and linked to Unity 3D~\cite{ Hussein2018ROSSimulation} and the Simulation of Urban Mobility (SUMO)~\cite{behrisch2011sumo} for a driver-centric and microscopic road traffic simulation, respectively. Furthermore, 3DCoAutoSim makes use of SUMO via the Vehicular Network Interface (Veins)~\cite{sommer2011bidirectionally} and is capable of analyzing and evaluating platooning mode from both a networking and a road traffic standpoint~\cite{validi2021simulation} by relying on the Platooning Extension for Veins, PLEXE~\cite{segata2014plexe}. Veins in turn uses the Objective Modular Network Test bed in C++ (OMNeT++)~\cite{OMNeT++Simulator} network simulator. 

In this work, we extended the platooning functionality of the 3DCoAutoSim simulation platform. Our system focused on opportunities for vehicles to rebuild a platoon after it fragmented by defining the behavior for the vehicles, thereby enabling spontaneous reformation. To validate our approach we relied on real driving data from daily trips obtained from
the Achleitner Organic farm in Upper Austria~\cite{Krawinkler2021}. We created from these trips a route as potential path to deliver Personal Protective Equipment (PPE) from China~\cite{kashansky2021adapt} to Linz.
For this purpose, we focused on several use cases for three electric vans with an autonomous (without driver) leader, a semi-autonomous first follower with a driver, and an autonomous last follower. Accordingly, we measured the relationship between the independent variables speed, acceleration and longitudinal gap and the dependent variable travel time in manual driving and dynamic-flexible platooning.
The remaining parts of this paper are organized as follows: the next section presents related work regarding the effects of electric vehicle platooning on traffic efficiency. Section~\ref{sec:ImplementationApproach} describes the methodology adopted to link the different modular components within the 3DCoAutoSim simulation platform and to implement the proposed dynamic-flexible platooning. The setup for the experiments and data acquisition and analysis are described in~\ref{sec:experimentandData}. The validation of the proposed model is presented in~\ref{sec:statistical}. The results from the analyses are presented in section~\ref{sec:Results}. The last section,~\ref{sec:conclusionfuturework}, discusses the findings, closes the paper and proposes future research.

\section{Related Work}
\label{sec:RelatedWork}

Several related challenges arise in semi-autonomous dynamic-flexible platooning when confronting the vehicles’ disconnection and reconnection to reform the platoon, a large issue being the maintenance of a constant longitudinal gap size among the platoon members. In this section, we review traffic efficiency when dealing with electric vehicles in platooning. 

The authors of~\cite{schellenberg2015electric} discussed the potential applications of wireless vehicular communication for electric vehicles in platooning mode to improve the battery life and usage. They reported several benefits that included reduced waiting time and shorter queue length at charging stations.

An improved Platoon Based (PB) multi-agent intersection management system simulated in SUMO was introduced in~\cite{Jin2013Platoon-basedVehicle}. Factors such as average time, pollutant emissions and message communication congestion were considered to analyze two use cases in which intersections were controlled with a traffic light control (TLC) and a Non-Platoon-Based (NPB) multi-agent intersection management system, respectively. An increase in environmental benefits of 20\% and an average travel time shortened by 30\% were reported from the implementation of the PB multi-agent intersection management system. According to the authors, the proposed system could also reduce the communication loads of an intersection management agent by up to 90\%.

A further Platoon-based experiment was performed in Korea by simulating a highway scenario to evaluate potential benefits in terms of travel time and congestion reduction~\cite{jo2019benefits}. To this end, a framework for macro and micro level simulation was adopted based on the microscopic traffic simulation  Verkehr  In  Stadten SIMulationsmodell (VISSIM)~\cite{VISSIM} and the macroscopic simulation transportation planning software (Trans-CAD)~\cite{TransCAD}. 
The micro-level analysis identified the impact of truck platoons on highway capacity. The macro-level analysis made it possible to assess the difference in travel times between scenarios comparing platooning mode with no platooning mode. Results from the 160 tested scenarios showed a $ 15\% $ travel time reduction on all Korean freeway networks when the vehicles were connected in platooning mode. 

Review of relevant literature on platooning shows that most of the analyses have been performed in highways, which is referred to as consistent platooning. Worth mentioning is also the fact that in all these studies only fuel combustion engines have been considered. The available literature on investigating dynamic (sometimes referred to as flexible) platooning and the approaches to how to manage, maintain and reform a platoon after it momentarily fragments as well as any potential effects on semi-autonomous electric vehicles is very scarce. Similarly, studies that use data generated from real road trips are very limited. 
Considering these gaps in literature, we contribute to the body of knowledge by developing, implementing and validating a dynamic-flexible platooning algorithm for electric vehicles as described in the next sections. 


\section{Implementation Approach}
\label{sec:ImplementationApproach}
To create the dynamic-flexible model we extended the platooning functionality of the 3DCoAutoSim simulation platform by adhering to
the SUMO platooning plugin Simpla~\cite{simpla2021}.
We then analyzed the effects of dynamic-flexible platooning on travel time compared to manual driving. To this end, we defined the travel time as dependent and speed, acceleration and longitudinal gap size as  independent variables. 
We considered three electric vans: one leader and two followers. We defined and implemented the leader and the last follower as autonomous electric vans without a driver, and the first follower as a semi-autonomous electric van, meaning that it could be driving in connected mode or manual mode. To validate our implementation process and to determine whether a statistically significant relationship existed between the defined variables, we performed a regression analysis with an F-test of the joint equality of the means and variances for the experimental and simulated measurements.
 
\subsection{Use Cases and Scenarios Definitions}
\label{sec:usecasescenario}

We developed two use cases and measured the gathered data in each of them through the following two modules of the 3DCoAutoSim simulator:
\begin{enumerate}
	\item Experimental data (exp) from real drivers in a designed route within the driver-centric module. Here the participants controlled the previously mentioned semi-autonomous electric delivery van that followed a leading van driving in autonomous mode. A third autonomous van followed the manually operated van.
	\item Simulated traffic behavior (sim) by feeding GPS-based data and further parameters that were validated in previous literature (see Table~\ref{table:simulationparameters}), into the microscopic traffic simulation environment relying on SUMO. 

\end{enumerate}

We defined the two use cases as follows:

\begin{table}
	\centering
	\small
	\caption{Parameter values to setup the use case simulations (sim)}
	\label{table:simulationparameters}
	\renewcommand{\arraystretch}{1,15}
	\resizebox{\linewidth}{!}{%
		\begin{tabular}{|c|l|l|c|} 
			\hline
			\multicolumn{2}{|c|}{\textbf{Parameters}} & \multicolumn{1}{c|}{\textbf{Value}} & \multicolumn{1}{l|}{\textbf{\textbf{~~~~~~~~~~~~~Source}}} \\ 
			\hline
			\multirow{2}{*}{Speed} & min & \multicolumn{1}{c|}{$5~ m/s (18~km/h)$} & \multirow{7}{*}{\begin{tabular}[c]{@{}c@{}}\textit{GPS}\\\textit{(electric delivery vans in Linz)}\end{tabular}} \\ 
			\cline{2-3}
			& max & \multicolumn{1}{c|}{$20~ m/s (72~km/h)$} &  \\ 
			\cline{1-3}
			\multicolumn{2}{|c|}{Acceleration (max)} & \multicolumn{1}{c|}{$2.5~m/s^2$} &  \\ 
			\cline{1-3}
			\multirow{2}{*}{Delivery van} & weight & $3500~kg$ &  \\ 
			\cline{2-3}
			& length & $5.94~m$ &  \\ 
			\cline{1-3}
			\multirow{2}{*}{Battery device} & capacity & $41000~Wh$ &  \\ 
			\cline{2-3}
			& power & $7200~Wh$ &  \\ 
			\hline
			\multicolumn{1}{|l|}{\multirow{2}{*}{Longitudinal gap}} & min~ & $7~m$ & \multirow{2}{*}{\textit{Adopted from~\cite{validi2021simulation}}} \\ 
			\cline{2-3}
			\multicolumn{1}{|l|}{} & max & $10~m$ &  \\
			\hline
		\end{tabular}
	}
\end{table}

\begin{figure}[!t]
	\centering
	\includegraphics[width=0.48\textwidth]{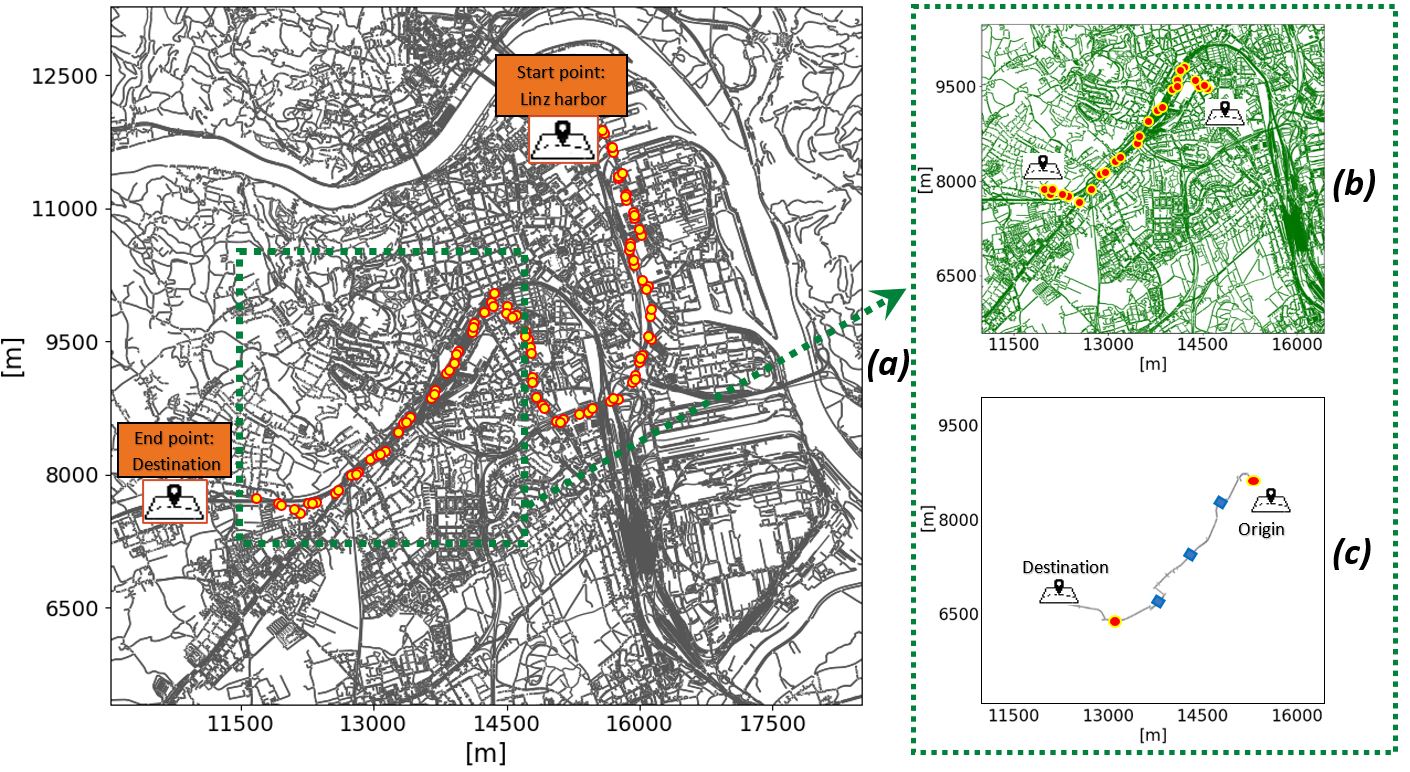}
	\caption{Traffic light positions on the main route (a) and the considered route selected part of the main route) (b). Visualization of the origin, the destination of the selected route, and the location on the map in which the platoon fragmentation occurred (the circles represent the position of the traffic lights that caused the disconnection, the squares represent other causes for disconnection such as other vehicles in close proximity) (c). } 
	\label{fig:tlsdisconnectionprofile}
\end{figure}

\begin{enumerate} [label=\Alph*)]
		\item Manual. The scenario consisted of three electric vans driving along a defined route according to the experimental data (exp) and simulated traffic behavior (sim).
	\begin{itemize}
		\item {(exp): the first follower (second van) was controlled by the driver during the whole trip.}
		\item {(sim): simulated traffic behavior of all the three vans completely controlled by the SUMO module. }
		
	\end{itemize}

	\item Dynamic-flexible platooning. Data was gathered in this use case using connected and non-connected vehicles. 
	\begin{itemize}
		\item {(exp): all three vans were connected. A constant connection between the leading van and the semi-autonomous first follower was not guaranteed, i.e. could be lost. As a consequence the driver needed to take over control to reconnect and restore the platoon}. 
		\item {(sim): simulated driving behavior. No driver involved.}
		
	\end{itemize}
\end{enumerate} 

nt platooning (B) vs Dynamic-flexible platooning (C)	

The defined route and the location in the map in which platoon fragmentation occurred are illustrated in Figure~\ref{fig:tlsdisconnectionprofile}. To evaluate the impact of platooning according to each of the defined use cases, we compared the measured parameters.

\subsection{Traffic Simulation Implementation}
\label{sec:simimplimentation}

To implement the use cases we relied on the representation of real data gathered from daily trips from electric delivery vans in Upper Austria into the SUMO module of the 3DCoAutoSim simulation platform. After importing the related map data we visualized the network in the SUMO Graphical User Interface (GUI). 
The resulting network was then updated with the corresponding traffic lights locations (Figure~\ref{fig:tlsdisconnectionprofile}). To replicate realistic traffic conditions, we applied real statistics data~\cite{Linzstatistics} as traffic demand using Iterative Assignment (Dynamic User Equilibrium) in three iterations (Activity Based Demand Generation). The data was then processed using $ OD2TRIPS $ to define the start and end point of a path; the route being then calculated in combination with $ DUAROUTER $. To mimic a dynamic traffic flow, speed and vehicle parameters ($ vType $), we calibrated the traffic flow in the simulation. After this, the values of the simulated objects could be extracted using TraCI. 

\subsection{Platooning Implementation}
We assumed in our approach that an appropriate V2V communication protocol existed, so this is not described in this work.
We adhered to the TraCI Python client, as it includes the Simpla configurable platooning plugin that enables application of several behavioral levels within the dynamic-flexible platooning model. 
We defined specific parameters such as maximum platoon gap and split time in a $ config.xml $ file.   
 
We additionally implemented the algorithm~\ref{algo:trigger} that generated the corresponding platoon behavior depending on the traffic conditions. The $simulation\_start$ function launched the SUMO application according to the implemented steps mentioned in section~\ref{sec:simimplimentation}. The $activate\_simpla\_extension$ parsed the $.xml $ file that included the platoon’s definition to be applied to the vans. The $if$ statements handled the relevant platooning model. The semi-autonomous electric delivery van was first verified in the algorithm. When it crossed a specified road that was managed by a defined traffic light, an event was triggered to apply the relevant disconnection event. Then, a message was sent to the driver to take over control of the semi-autonomous electric delivery van and approach the leading vehicle to reconnect and reform the platoon.

\begin{algorithm}[!t]
	\caption{\small Dynamic-flexible platooning}
	\label{algo:trigger}
	\small	
	\SetKwData{Left}{left}\SetKwData{This}{this}\SetKwData{Up}{up}
	\SetKwComment{Comment}{/* }{ */}
	\SetKwFunction{Union}{Union}\SetKwFunction{FindCompress}{FindCompress}
	\SetKwInOut{Input}{input}\SetKwInOut{Output}{output}\SetKwInOut{Define}{define}
	\Input{ $traffic\_net$; $tID$; $lID$; $eID$}
	\Output{$dynamic\_model$}
	\BlankLine
	$simulation\_start()$\;
	$leader\_edge\_id \gets get\_road\_id(lID)$\;
	$time \gets 10$\Comment*[r]{10 sec} 
	\BlankLine
	\If{$platoon\_mode$ == $"dynamic"$}{
		$activate\_simpla\_extension()$\;
	
	\BlankLine
	\If{$leader\_edge\_id$ == $eID$}{
		\BlankLine
		$setTrafficLight(tID, "RED", time)$\;
		$wait\_while\_leader\_is\_connected()$\;
		$send\_alarm\_to\_player()$\;
		\BlankLine
	}
	\BlankLine
	\While{$getMinExpectedNumber() > 0$}{
		\BlankLine
		\If{$platoon\_mode$ == $"dynamic"$}{
			$break\_platoon(tID, lID)$\;
		}
		\BlankLine
	}
}
\end{algorithm}


\section{Experimental Setup and Data Analysis}
\label{sec:experimentandData}

We set the simulation time step to 0.01 seconds for all of the experiments, implying a 20Hz update frequency for the driver-centric simulator and SUMO.
Prior to conducting the experiments in the simulator the participants were provided with detailed information regarding the operation and experiment development, including the data privacy policy. The participants were asked to adhere to all applicable traffic laws from the origin to the destination.

\begin{figure}[!t]
	\label{fig:experimnetmanual}
	\centering
	\includegraphics[width=0.50\textwidth]{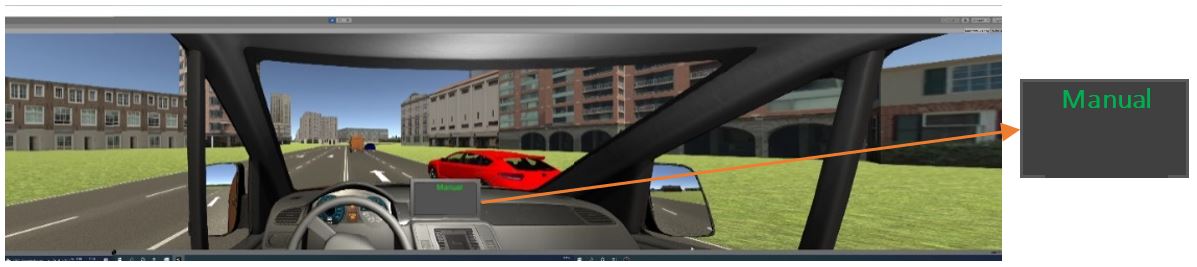}
	\caption{Visualization of driving experiment. The in-vehicle display shows the driving mode as manual (use case A).}
\end{figure}

The sample consisted of $ 17 $ participants ($ 15 $ males and $ 2 $ females, mean age = 49.7 (between 24 to 52). We divided the experiments into two sets as follows:

\begin{enumerate}
	\item Manual driving (exp): participants drove the first follower (middle electric delivery van):
	\begin{itemize}
		\item from origin to destination,
		\item trying to maintain a constant longitudinal gap size from the first van, equal to $ 7m $.
	\end{itemize}
	\item Dynamic-flexible platooning driving (exp): participants driving the first follower vehicle, performed the driving task in connected, platooning mode until a reduction of speed caused by the location of a traffic light fragmented the platoon. Figure~\ref{fig:tlsdisconnectionprofile} (b) illustrates the points in which the platoon was dissolved. In order to visualize the status of the connection and the longitudinal gap size from the leading vehicle, we additionally equipped the vans with an in-vehicle system that showed the driving mode. Figure~\ref{fig:experimnetmanual} depicts the system in use case A (manual). To perform this experiment the participants were asked to: 
	\begin{itemize}
		\item take control of the vehicle in the event of a dissolving of the platoon (the in-vehicle display changed in this case from connected to manual and emitted an acoustic signal),
		\item try to maintain a longitudinal gap size from the leading vehicle of close $ 7m $,
		\item accelerate to reach the leading vehicle in the event of being disconnected. 
	\end{itemize}
\end{enumerate}

Speed and acceleration data from the semi-autonomous electric delivery van were gathered through the performed experiments within a time interval of $ 0,01 $ seconds. The longitudinal gap size and the travel time of the obtained data were then calculated to further proceed with the pre-processing, cleaning and analysis of the data set.

Within the process we applied the Sklearn’s Outlier detection with Local Outlier Factor (LOF)~\cite{lofoutlier} to locate and remove 0,05\% of the outliers in the data set, and thus be able to perform the subsequent validation process in a more accurate way.

\section{Validation of the Proposed Dynamic-Flexible Platooning Model}
\label{sec:statistical}

\begin{figure*}[!t]
	\centering	
	\includegraphics[width=.72\textwidth]{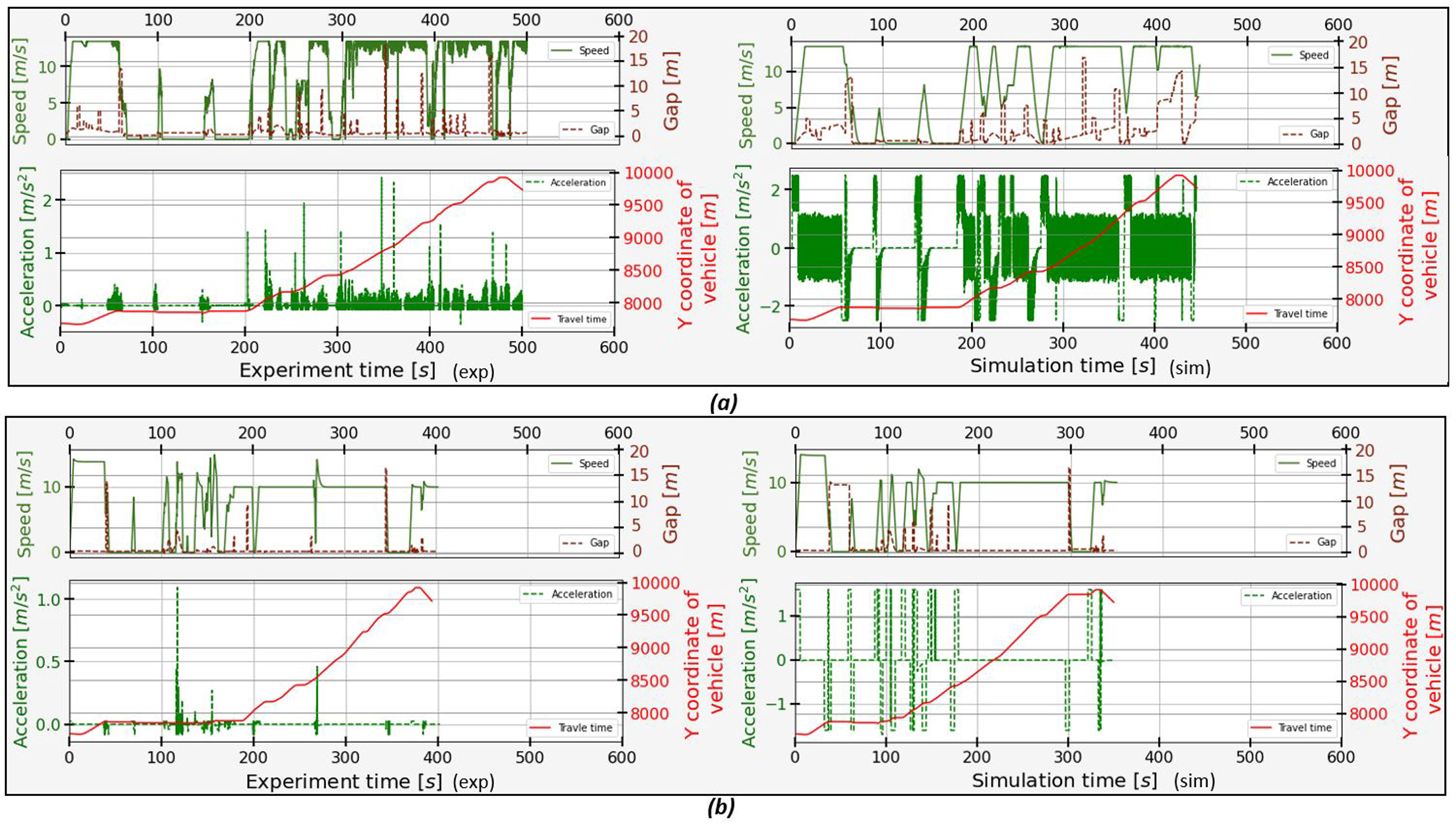}
	\caption{Illustration of the (exp) and (sim) module-related data for the semi-autonomous electric van in manual (use case A) (a) and dynamic-flexible platooning (use case B) (b)}
	\label{fig:comparativmanual}
\end{figure*}

\textcolor{black}{To determine the degree to which the simulation model and its associated data are an accurate representation of the real world, 
we conducted a regression analysis and applied the F-statistic analysis. For this purpose we relied on the approach in~\cite{toledo2004statistical}.} 
Our metamodels described the structure of the output data obtained from the simulations (sim) and the performed experiments with real drivers involved (exp) to calculate travel time. 

We tested the following working hypothesis $ H_1:\beta^{exp} = \beta^{sim} $ against the null hypothesis $ H_0:\beta^{exp}\neq\beta^{sim} $. Equation~\ref{eq:ess} shows the applied F-test, adopted from~\cite{toledo2004statistical}.

\begin{equation}
	\tag{1}\
	\label{eq:ess}
	F = \frac{(ESS^{R}-ESS^{UR}) \ / \ K}{ESS^{UR} \ / \ [N^{sim}+N^{exp}-2K]} 
\end{equation}

where $ ESS^{R} $ and $ ESS^{UR} $ are sums of squared residuals of the
restricted and the unrestricted models, respectively, calculated as  $ ESS^{R} $ = $ ESS^{com} $ and $ ESS^{UR} = ESS^{exp}+ ESS^{sim}$.
$ ESS^{com}, ESS^{exp}$, and $ESS^{sim} $ result from the statistical analysis of the combined, the observed, and the simulated data, respectively. $ N^{exp} $ and $ N^{sim} $ represent the number of experiments in the (exp) and (sim), respectively and
$ K $ denotes the number of parameters in the model. The base regression analysis model is: $ y  = \beta_0  +  \beta_1x  + \epsilon$,  where $ y $ denotes the dependent variable travel time, 
 $ \beta_0 $ the intercept, $ \beta_1 $ the independent variable coefficient, $ x $ denotes the independent variable (speed, acceleration or gap) and $ \epsilon $ denotes the estimated error.

\section{Analyses Results}
\label{sec:Results}

\subsection{Comparative Analysis of the Use Cases}
\label{subsec:ComparativeUseCases}

The effect of manually operated and dynamic-flexible platooning modes on travel time is presented in Figure~\ref{fig:comparativmanual}. The $ 450~sec $ travel time in manual driving (sim) was reduced by $ 23.77\% $ to $ 343~sec $ when the vehicles were connected in dynamic-flexible platooning. Driving in manual (exp) was longer with $ 508~ sec $ compared to $ 400~sec $ in driving in dynamic-flexible platooning. A travel time reduction of $ 21.25\% $ was observed when vehicles were in dynamic-flexible platooning mode. Results showed that the benefits of connection in terms of travel time persist even if the connection is disrupted for certain periods of platooning and the driver needs to regain control of the vehicle.  

\begin{figure*}[!t]
	\centering
	\includegraphics[width=.86\linewidth]{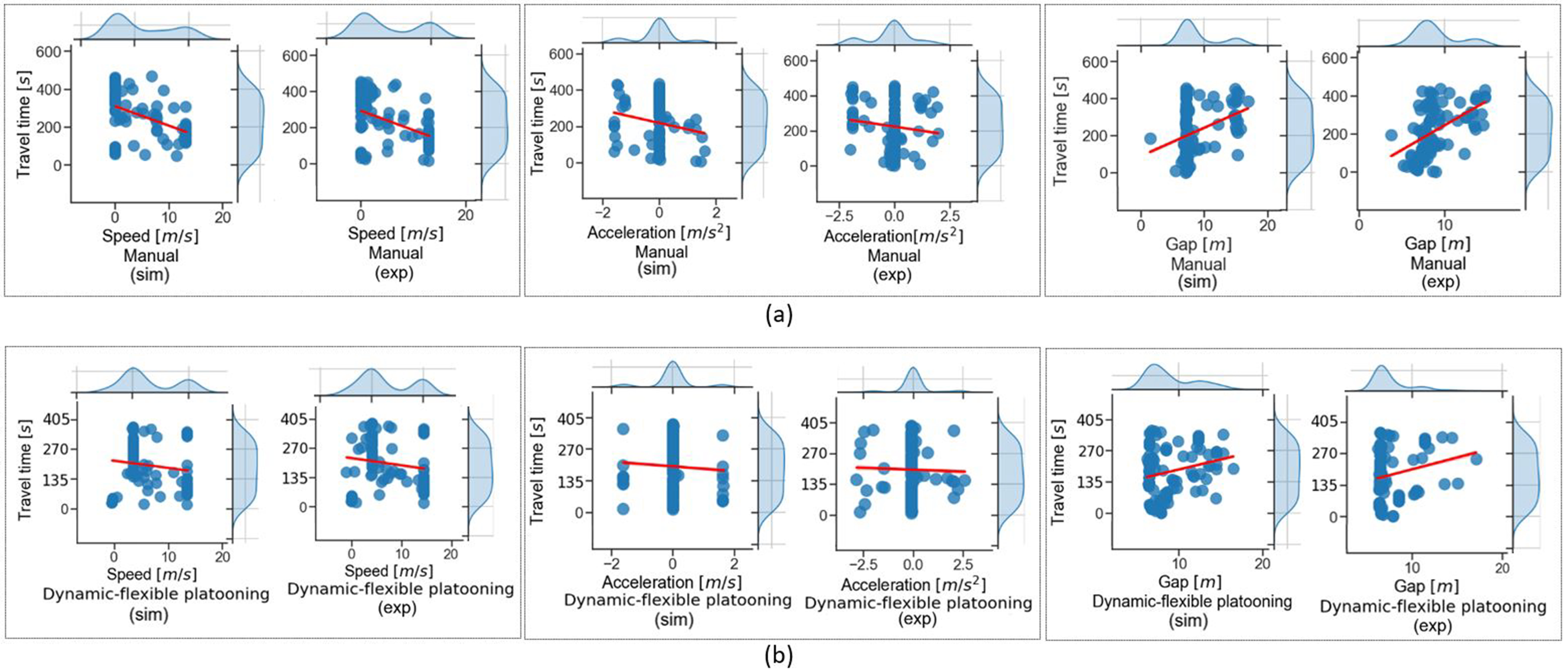}
	\caption{Simple linear regression results of the relationship between the data gathered through the experimental setup in the driver-centric module of the 3DCoAutoSim simulation platform (exp) and the data gathered by feeding the correspondent values into the traffic simulation module (sim). Each graphic visualizes the effect of the independent variables speed, acceleration and longitudinal gap on the dependent variable travel time). (a) represents manual driving and (b) dynamic-flexible platooning.}
	\label{fig:regressiontraveltime}
\end{figure*}

\begin{table*}
	\centering
	\caption{Regression models' results }
	\label{table:regressionresults}
	\renewcommand{\arraystretch}{1.1}
	\resizebox{\linewidth}{!}{%
		\begin{tabular}{|l|l|l|l|l|l|c|c|l|} 
			\hline
			\multicolumn{2}{|c|}{\multirow{2}{*}{\diagbox{\textbf{Independent variables}}{\begin{tabular}[c]{@{}l@{}}\textbf{Regression models}\\\textbf{details}\end{tabular}}}} & \multicolumn{4}{c|}{\textbf{Regression parameters}} & \multicolumn{2}{c|}{\textbf{Regression line}} & \multirow{2}{*}{\textit{\textbf{\textbf{\textbf{\textbf{F-Statistic}}}}}} \\ 
			\cline{3-8}
			\multicolumn{2}{|c|}{} & \textit{\textbf{\textbf{R-Squared}}} & \multicolumn{1}{c|}{\textbf{\textit{ESS}}} & \textbf{\textit{P-Value}} & \textbf{\textit{Mean}} & \begin{tabular}[c]{@{}c@{}}\textbf{\textit{Intercept}}\\\textbf{\textit{($\beta_0$}})\end{tabular} & \begin{tabular}[c]{@{}c@{}}\textbf{\textit{Independent variable }}\\\textbf{\textit{Coef ($\beta_1$)}}\end{tabular} &  \\ 
			\hline
			\multirow{2}{*}{\textbf{\textit{Manual~speed}}} & \textbf{\textbf{\textit{sim}}} & 0.50007 & 192784.80 & 0.00088*** & 10.26 & 331.51 & -11.86 & 17.915 \\
			& \textbf{\textbf{\textbf{\textbf{\textbf{\textbf{\textbf{\textbf{\textit{exp}}}}}}}}} & 0.50313 & 192783.25 & 0.00083*** & 11.13 & 331.78 & -11.86 & 16.874 \\ 
			\hline
			\multirow{2}{*}{\textbf{\textbf{\textit{Manual~acceleration}}}} & \textit{\textbf{sim}} & 0.49915 & 203645.47 & 0.00072*** & 0.012 & 302.58 & -10.31 & 18.251 \\
			& \textbf{\textit{exp}} & 0.50177 & 203644.89 & 0.00061*** & 0.019 & 301.34 & -9.01 & 18.986 \\ 
			\hline
			\multirow{2}{*}{\textbf{\textit{Manual~gap }}} & \textbf{\textit{sim }} & 0.49117 & 194165.87 & 0.00079*** & 10.05 & 111.26 & 12.32 & 17.786 \\
			& \textit{\textbf{exp}} & 0.49897 & 194164.32 & 0.00082*** & 10.68 & 110.34 & 12.01 & 17.783 \\ 
			\hline
			\multirow{2}{*}{\textbf{\textit{Dynamic-flexible platooning~speed}}} & \textbf{\textit{sim}} & 0.49899 & 248536.07 & 0.00077*** & 9.78 & 205.16 & -10.26 & 17.971 \\
			& \textbf{\textit{exp}} & 0.50827 & 248536.98 & 0.00080*** & 9.96 & 205.79 & -10.21 & 17.954 \\ 
			\hline
			\multirow{2}{*}{\textbf{\textit{Dynamic-flexible platooning~acceleration }}} & \textbf{\textbf{\textit{sim}}} & 0.49521 & 253179.22 & 0.00076*** & 0.025 & 200.54 & -7.24 & 19.387 \\
			& \textbf{\textit{exp}} & 0.50348 & 253180.10 & 0.00083*** & 0.033 & 200.02 & -7.01 & 19.387 \\ 
			\hline
			\multirow{2}{*}{\textbf{\textit{Dynamic-flexible platooning~gap }}} & \textbf{\textbf{\textit{sim}}} & 0.49329 & 240403.68 & 0.00062*** & 9.94 & 140.54 & 11.77 & 18.954 \\
			& \textbf{\textit{exp}} & 0.49998 & 240402.35 & 0.00075*** & 10.02 & 140.32 & 11.60 & 18.657 \\ 
			\hline
			\multicolumn{9}{l}{\textcolor{black}{*A high level of unexplained variation often exists in studies involving human behavior, thus the $ R^2 $  values are generally lower than $ 50\% $ }}
		\end{tabular}
	}
\end{table*}


\subsection{Regression Analysis}
\label{subsec:regressionanalysis}

The results of the regression models analysis for the use cases manual driving (A) and dynamic-flexible platooning (B) are depicted in Figure~\ref{fig:regressiontraveltime} 
that shows the regression lines for the dependent variable travel time and the independent variables speed, acceleration and longitudinal gap size respectively. They are the result of the relationship between the data gathered through the experiments in the driver-centric module of the 3DCoAutoSim simulation platform (exp) and the data gathered by feeding the correspondent values into the traffic simulation module (sim). 
The distribution of the data and the regression lines show that the simulation data set follows the same general pattern that exists in the experiment's data set. 

The two sets of related data speed-travel time and acceleration-travel time showed a negative correlation, as an increase in speed and acceleration occurred alongside a decrease in travel time, and a continued reduction in travel time could be expected in the time frame over the investigated period. The increase in travel time, due to the increase in gap size, is also reflected in the depicted relationship between the two data sets in the figure.

Table~\ref{table:regressionresults} presents the results from applying the regression models for the simulated (sim) and the experimental datasets (exp). R-Squared as a measure of goodness-of-fit, measures the strength of the relationship between travel time (dependent variable) and the independent variables speed, acceleration and gap. The values from the F-Statistic column show how well the regression equation explained the variation in the sim and exp data set. The ESS gives an estimate of how well the correspondent model explained and modeled the observed data. As result from the statistically significant relationship between the dependent and independent variables indicated by the P-values as well as the similar values in the $\beta_0$  and $\beta_1$ in both sim and exp, we reject the null hypothesis $ H_{0} : \beta^{exp} \neq\beta^{sim}$ and accept $ H_{1} :\beta^{exp} = \beta^{sim}$.\\

\section{Conclusion and Future Work}
\label{sec:conclusionfuturework}

In this work we investigated the impact of dynamic-flexible platooning on travel time. To this end, we extended the platooning capabilities of the 3DCoAutosim simulation platform.  The regression analysis, has been proven to be an appropriate approach that provides sufficient evidence (R-Squared, $ ESS $, P-Value) to validate the model. As previously mentioned, R-squared measures the strength of the relationship between the model and the independent variables. Generally, the closer the R-squared value is to $ 100\% $ the stronger the relationship. However, in studies involving human behavior, the R-squared values are generally lower than $ 50\% $~\cite{meyer2013semi}. In some disciplines a higher level of unexplained variation simply quite often exists due to human behavior. Therefore we conclude that the low R-squared values presented in Table~\ref{table:regressionresults}, are valid and are due to the unpredictability of human behavior. The dataset from the performed experiments with drivers in the driver-centric simulation module as well as the data from feeding the pertinent values in the traffic simulator module with the same dependent and independent variables showed a similar trend. These results validate our presented approach in this paper. We conclude that the model developed in this study is proven to be efficient to investigate the effects of dynamic-flexible platooning on travel time. Future work will extend our proposed dynamic-flexible platooning model by considering vehicles entering and/or leaving the platoon. 


\section*{Acknowledgment}
This work was partially supported by the ''ADaptive and Autonomous data Performance connectivity and decentralized Transport decision-making Network (ADAPT)'' project funded by the Austrian Research Promotion Agency (grant agreement 881703) and the Chinese Academy of Science; the FFG ''Zero Emission Roll Out Cold Chain Distribution\_877493'' project and the
Austrian Ministry for Climate Action, Environment, Energy,
Mobility, Innovation and Technology (BMK) Endowed Professorship for Sustainable Transport Logistics 4.0., IAV France
S.A.S.U., IAV GmbH, Austrian Post AG and the UAS Technikum Wien.



\bibliographystyle{IEEEtran}
\bibliography{references}

\end{document}